# Versatile tunable optical injection of chiral polarized weyl fermions in a magnetic weyl semimetal $Co_3Sn_2S_2$

Zipu Fan[1], Junchao Ma[1], Jinying Yang[2,3], Yan Sun[4], Zhuocheng Lu[1], Shuxia Chen[1], Delang Liang[1,5], Dehong Yang[1], Chang Xu[1], Qinsheng Wang[6], Anlian Pan[5], Ji Feng[1], Enke Liu[2,*], JinLuo Cheng[7,*], Dong Sun[1,8,9,*]

[1]International Center for Quantum Materials, School of Physics, Peking University, Beijing 100871, China
[2]Beijing National Laboratory for Condensed Matter Physics, Institute of Physics, Chinese Academy of Sciences, Beijing 100190, China
[3]School of Physical Sciences, University of Chinese Academy of Sciences, Beijing 100190, China
[4]Shenyang National Laboratory for Materials Science, Institute of Metal Research, Chinese Academy of Sciences, Shenyang 110016, China
[5]Key Laboratory for Micro-Nano Physics and Technology of Hunan Province, Hunan Institute of Optoelectronic Integration, College of Materials Science and Engineering, Hunan University, Changsha 410082, China
[6]Key Laboratory of Advanced Optoelectronic Quantum Architecture and Measurement (Ministry of Education), School of Physics, Beijing Institute of Technology, Beijing 100081, China
[7]GPL Photonics Laboratory, State Key Laboratory of Luminescence Science and Applications, Changchun Institute of Optics, Fine Mechanics and Physics, Chinese Academy of Sciences, Changchun 130033, China
[8]Collaborative Innovation Center of Quantum Matter, Beijing 100871, China
[9]Frontiers Science Center for Nano-optoelectronics, School of Physics, Peking University, Beijing 100871, China
*Corresponding authors. E-mails: ekliu@iphy.ac.cn; jlcheng@ciomp.ac.cn; sundong@pku.edu.cn

## ABSTRACT

Precise probe and control of various quantum degrees of freedom in novel quantum matter are central to understanding fundamental quantum physics and hold promise for innovative routes to encode and process information. Chirality is one such degree of freedom that has recently attracted intense research interest, especially for Weyl fermions in topological Weyl semimetals. The coupling of chiral degrees of freedom through light–matter interactions and the versatile control of these couplings through external fields can lead to precise quantum control of Weyl fermions. In this work, we demonstrate the observation of light chirality-dependent photocurrent in the mid-infrared regime. Excitation wavelength-dependent measurements reveal that the photocurrent originates from the injection of chiral polarized Weyl fermions by chiral polarized mid-infrared photons. The optical process that generates unbalanced chiral polarized Weyl fermions is determined to be a third-order nonlinear photocurrent process. Compared with nonmagnetic Weyl semimetals, such coupling is versatilely tunable in magnetic Weyl semimetals with the magnetization direction and external electric field in addition to the chirality of light. Our results are not only directly applicable to tunable circular-polarization-sensitive photodetection in the mid-infrared regime, but also pave the way toward functional quantum devices that utilize the chiral quantum degrees of freedom of Weyl fermions.

# INTRODUCTION

Chirality typically tells whether a physical entity, such as a particle, is equivalent to its mirror image. For massless particles, chirality equals helicity, which is a more intuitive concept and can be understood as the projection of the spin in the direction of motion. A familiar example is circularly polarized light whose winding direction follows either the right or left hand, which corresponds to opposite chirality or is mostly termed "helicity" for photons. In Weyl semimetals [1–3] (WSMs), the chirality of the Weyl fermions, which is equivalent to the helicity of a massless particle, refers to whether the directions of spin and motion are parallel or antiparallel and is determined by the sign of the monopole charge of the Weyl nodes. A right-handed Weyl node ($\chi_W = +1$) and a left-handed Weyl node ($\chi_W = -1$) are the monopole and antimonopole of Berry curvature and carry Chern numbers C = +1 and −1, respectively. The coupling between the polarization chirality of light and the chirality of a Weyl cone is determined by the selection rules of the optical transitions in the Weyl cone [4, 5] (Fig. 1a). The absorption of a circularly polarized photon flips the electron spin to conserve the total angular momentum of the photon and electron: the optical transitions from spin-up to spin-down states require left circularly polarized (LCP) excitation, whereas those from spin-down to spin-up states require right circularly polarized (RCP) excitation. Here, we note that the two bands around a Weyl cone in Weyl semimetals can also be formed by pseudospins, which can provide similar optical selection rules as the spin bands. For simplicity, we still use the terminology "spin" for the Weyl bands. Thus, photons with different circular chiralities are absorbed at opposite sides of the Weyl cone and generate carriers with opposite momentum directions. Consequently, the circularly polarized light will generate a directional electric current in the Weyl cone, with the current direction governed by the chirality of the Weyl cone. However, such a scenario is not trivial to realize experimentally in Weyl semimetals, as Weyl cones always come in pairs with opposite chirality, and the electric current generated by the injected chiral polarized Weyl fermions normally cancel out in a pair of Weyl cones with opposite chirality [4]. Experimentally, it is also difficult to distinguish the response of the chiral polarized Weyl fermions from the contribution of carriers in trivial bands and other unrelated circular light polarization-dependent responses.

Chiral couplings between chiral polarized light and Weyl fermions and the injection of chiral polarized Weyl fermions via chiral polarized light have been realized in several IS-breaking WSMs, such as TaAs [5, 6] and TaIrTe$_4$ [7, 8], and mirror-symmetry breaking chiral semimetals RhSi [9, 10] and CoSi [11], benefiting from the nonzero second-order nonlinear response in IS-breaking materials. Chiral polarized light can create chiral polarization of Weyl fermions in IS-breaking materials when the Weyl cone has a finite tilt to avoid the generated photocurrent cancellation of opposite contributions from a pair of Weyl cones with opposite chiralities [4]. Weyl fermions can exist in materials that break either the inversion symmetry (IS) [3, 12–14] or the time-reversal symmetry (TRS) [15–17]. In TRS-breaking WSMs with preserved IS, the second-order nonlinear photocurrent response vanishes, and a previous strategy to generate chiral polarized Weyl fermions in IS-breaking Weyl semimetals is no longer applicable. Nevertheless, the inversion symmetry can still be broken by applying a static electric field, thereby enabling the generation of a photocurrent response through a third-order nonlinear optical process [18, 19]. Such third-order nonlinear photocurrent has also been widely observed in many other materials [7, 20, 21]. In this work, by applying a static electric field, we demonstrated that a nonvanishing third-order nonlinear photocurrent response can be triggered, and with the excitation of a chiral polarized mid-infrared photon, unbalanced chiral Weyl fermions can be injected in a TRS-breaking Weyl semimetal.

This work was performed on $Co_3Sn_2S_2$, a well-established ferromagnetic (FM) Weyl semimetal that has many exotic physical phenomena related to its topological Weyl cone, such as its unusual large anomalous Hall conductivity [22, 23] and giant magneto-optical response [24]. In $Co_3Sn_2S_2$, the easy axis of the magnetization lies along the out-of-plane direction [25, 26]. The chirality of the Weyl cone is directly related to the FM order and can be controlled externally by switching the magnetization direction [27], which provides an ideal platform for studying the chirality coupling between the excitation light and magnetic Weyl cones (Fig. 1b). Specifically, we perform circular-polarization-dependent photocurrent measurements to study the chirality coupling between the excitation light and magnetic Weyl cones in $Co_3Sn_2S_2$. In the mid-infrared wavelength range, our results show that $Co_3Sn_2S_2$ exhibits a significant light chirality-dependent photocurrent in the FM phase when an external in-plane electric field is applied. Further wavelength-dependent measurements revealed that the light chirality-dependent photocurrent only occurs under low-energy mid-infrared photon excitation, which suggests a dominant contribution from the couplings with Weyl cones. In addition, we also observe a sign switch of the chirality-dependent photocurrent at different mid-infrared wavelengths, which corresponds to the unbalanced injection of chiral polarized Weyl fermions at opposite sides of Weyl cones through chirality couplings between light cones and magnetic Weyl cones. Further numerical analysis helps determine that the contributions from topologically trivial magnetic circular dichroism are minor, which indicates that the observed light chirality-dependent photocurrent generation is dominated by the electric current generated from chiral polarized Weyl fermions injected by chiral mid-infrared photons. With flexibly tunable chiral polarized Weyl fermion injection, our work establishes magnetic Weyl semimetals as more versatile material platforms than IS-breaking Weyl semimetals for chirality manipulation and lays the basis for future quantum devices on the basis of chiral degrees of freedom.

## RESULTS

### Basic characterization of $Co_3Sn_2S_2$

$Co_3Sn_2S_2$ is an FM crystal with a Curie temperature of $T_C \sim 177$ K [28]. The crystal structure of $Co_3Sn_2S_2$ is shown in (Fig. 1c), and it belongs to the R-3m (no.166) space group and $D_{3d}$ point group. The magnetic cobalt atoms form a kagome lattice with a magnetic moment of 0.29 $\mu_B$/Co$^{?+}$ [28]. The magnetic easy axis is perpendicular to the kagome planes and along the $c$ axis [25, 26]. The magnetization direction naturally defines two opposite chiralities following the spin direction of the out-of-plane FM order. When the temperature falls below the $T_C$, the out-of-plane FM order breaks the TRS, and $Co_3Sn_2S_2$ enters the topological Weyl semimetal state [16, 29]. There are three pairs of Weyl points within each bulk Brillouin zone, which are 60 meV above the Fermi level according to ab initio calculations [22, 30]. In addition to the changes in the band structure, the magnetization of Co atoms also restricts the symmetry operation and lowers the symmetry of the system to $S_6$.

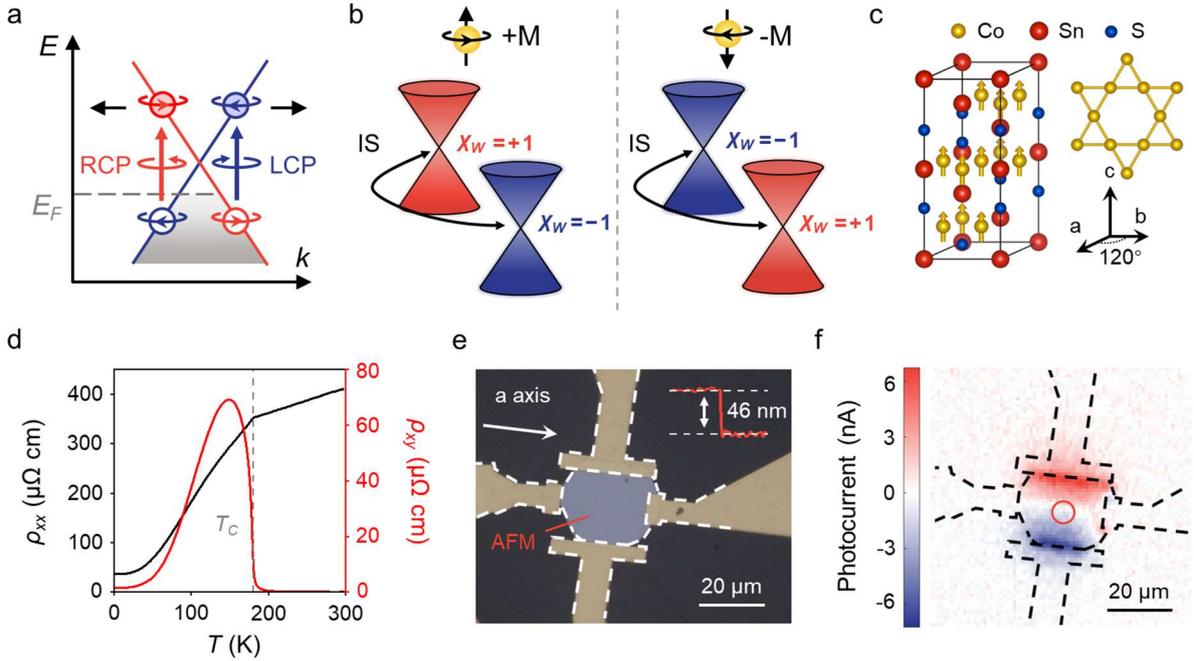

**Figure 1.** Basic characterization of $Co_3Sn_2S_2$. (a) Optical selection rules in the vicinity of a Weyl cone. The gray dashed line marks the Fermi level. (b) Schematic diagram of the relationship between the chirality of the Weyl cone and the FM order. The pair of Weyl cones with opposite chirality is related by inversion symmetry. (c) Crystal structure of $Co_3Sn_2S_2$. The magnetic moments are shown along the c-axis. The kagome lattice formed by the cobalt atoms is also shown. (d) Temperature dependence of the longitudinal ($\rho_{xx}$) and transverse ($\rho_{yx}$) resistivities under a magnetic field of 0.1 T along the *c*-axis. (e) Optical microscopy image of $Co_3Sn_2S_2$ device 1. The inset shows the thickness of $Co_3Sn_2S_2$ flake (~46 nm) measured by atomic force microscopy (AFM). The AFM scanning was performed along the red line. (f) Scanning photocurrent microscopy image of device 1 under 4-μm excitation. The measurement was taken after the sample was cooled to 140 K in a field of −0.25 T. The excitation power was 1.5 mW. The red circle denotes the beamspot position where the following experiments are performed.

The $Co_3Sn_2S_2$ flakes used in this work were grown via a chemical vapor transport method (see Appendix for details). Figure 1d shows the typical electrical transport measurement results for the $Co_3Sn_2S_2$ flakes. For electrical measurements, a $Co_3Sn_2S_2$ flake was fabricated into a standard Hall bar device, and the longitudinal and transverse resistivities were measured. The results show ferromagnetism with a Curie temperature close to that previously reported [31]. For optical measurements, the sample (~ 46 nm thick) was fabricated into a four-electrode device (device 1) with two pairs of electrodes arranged along the *x*-axis and the *y*-axis (Fig. 1e). Here, we establish the experimental coordinates as follows: the *x*-axis and the *z*-axis are oriented along the *a*-axis and *c*-axis of the crystal, respectively, and the *y*-axis is oriented perpendicularly to the *a*-axis. As a general characterization, we first perform scanning photocurrent microscopy (SPCM) measurements to determine the spatial distribution of the photocurrent. The photocurrent $I_{ph}$ is measured after the sample is cooled to 140 K in a magnetic field of −0.25 T (see Appendix for details), and the magnetization of the sample is denoted as -*M*. Figure 1f shows a typical SPCM image of device 1 under 4-μm excitation. The photocurrent is collected under unbiased mode using a pair of electrodes along the *y*-axis, while the other electrodes are floated. The SPCM image shows that the photocurrent is limited to the edges of the electrodes and that the current directions are opposite (opposite signs) for the two opposite electrodes. The lack of photocurrent in the central area of the sample is consistent with the fact that $Co_3Sn_2S_2$ possesses inversion symmetry and thus has a vanishing second-order nonlinear photocurrent response.

**Tunable light chirality-dependent photocurrent in $Co_3Sn_2S_2$**

Although the second-order nonlinear photocurrent response vanishes because of the IS of $Co_3Sn_2S_2$, the photocurrent can be induced through symmetry-breaking mechanisms, such as the presence of electrodes or, as focused in our work, by applying a static electric field. With the assistance of a static electric field, the photocurrent generated by the third-order nonlinear process can be expressed as:

$$I_a^{ph} = \sigma_{abcd} E_b^{dc} E_c^l (E_d^l)^*, \tag{1}$$

where $I_a^{ph}$ is the photocurrent, $\sigma_{abcd}$ is the third-order nonlinear coefficient, $E_b^{dc}$ is the static electric field provided either by the built-in field or the external bias field, $E_c^l$ is the light electric field provided by the optical excitation, and the subscript indicates the direction of the corresponding vector. Here, we use $\sigma_{abcd}$ to represent the contributions of all third-order nonlinear effects without distinguishing the specific microscopic mechanisms behind the photocurrent, which is unnecessary for analysis at this stage. Taking the photocurrent collected along the *y*-axis as an example, the FM phase (the $S_6$ point group) has a contribution from nonzero third-order nonlinear coefficients $\sigma_{yyyy}$, $\sigma_{yyxx}$, $\sigma_{yxxy}$, $\sigma_{yxyx}$, $\sigma_{yyyx}$, $\sigma_{yyxy}$, $\sigma_{yxyy}$, $\sigma_{yxxx}$. Here, we consider only the in-plane nonzero third-order nonlinear coefficients according to our measurement geometry, where the excitation light is normally incident along the *z* direction and the photocurrent is collected within the *x*–*y* plane. Therefore, a nonzero photocurrent along the *y*-axis can be induced by applying an external bias voltage along the *y*-axis, which is contributed by the nonzero third-order nonlinear coefficients $\sigma_{yyyy}$, $\sigma_{yyxx}$, $\sigma_{yyyx}$, $\sigma_{yyxy}$.

According to the symmetry analysis above, the nonzero photocurrent at the electrode interface relies on the symmetry breaking due to the electrode, such as the built-in electric field formed at the interface [32] or the different thermoelectric coefficients through photo-thermal effect [33]. In the studies presented below, we focus on the photocurrent response at the central area of the sample (as denoted by the red circle in Fig. 1f), where the photoresponse vanishes when no external bias is applied, to minimize the influence of the electrode interface. In the measurement, an external bias voltage is applied to provide a controllable $E_b^{dc}$ term in Equation (1). To study the chirality coupling between the excitation light and magnetic Weyl cones, we performed light chirality-dependent photocurrent measurements and the experimental scheme is shown in Fig. 2a. The chirality of the light is controlled by rotating a quarter wave plate (QWP), and the photocurrent is recorded as a function of the rotation angle of the QWP ($\theta_{\lambda/4}$). As shown in Fig. 2b, when a 5-mV external bias voltage is applied, the magnitude of the photocurrent is different under optical excitation with different chiralities (LCP for 45° and RCP for 135°), which indicates the existence of a light chirality-dependent photocurrent $I_A$ (the sign of $I_A$ is defined as positive if the $I_{RCP}-I_{LCP} > 0$). In contrast, no clear $I_A$ is observed in the absence of an external bias, as shown in Fig. 2b. To confirm that $I_A$ is indeed due to a third-order nonlinear optical response, we examined the dependence of the photocurrent on the light intensity by carrying out excitation power-dependent photocurrent measurements. As shown in Fig. 2c, both the total photocurrent and the $I_A$ increase monotonically with increasing excitation power. For clarity, we extract the polarization-independent component $I_{dc}$ and the light chirality-dependent component $I_A$ through Fourier transform and plot them as functions of the excitation power in Fig. 2d (details for the Fourier transform can be found in Supplementary Material Section I). Both the magnitudes of $I_{dc}$ and $I_A$ exhibit a linear dependence on the excitation power, implying a second-order dependence on the light electric field, which is consistent with Equation (1).

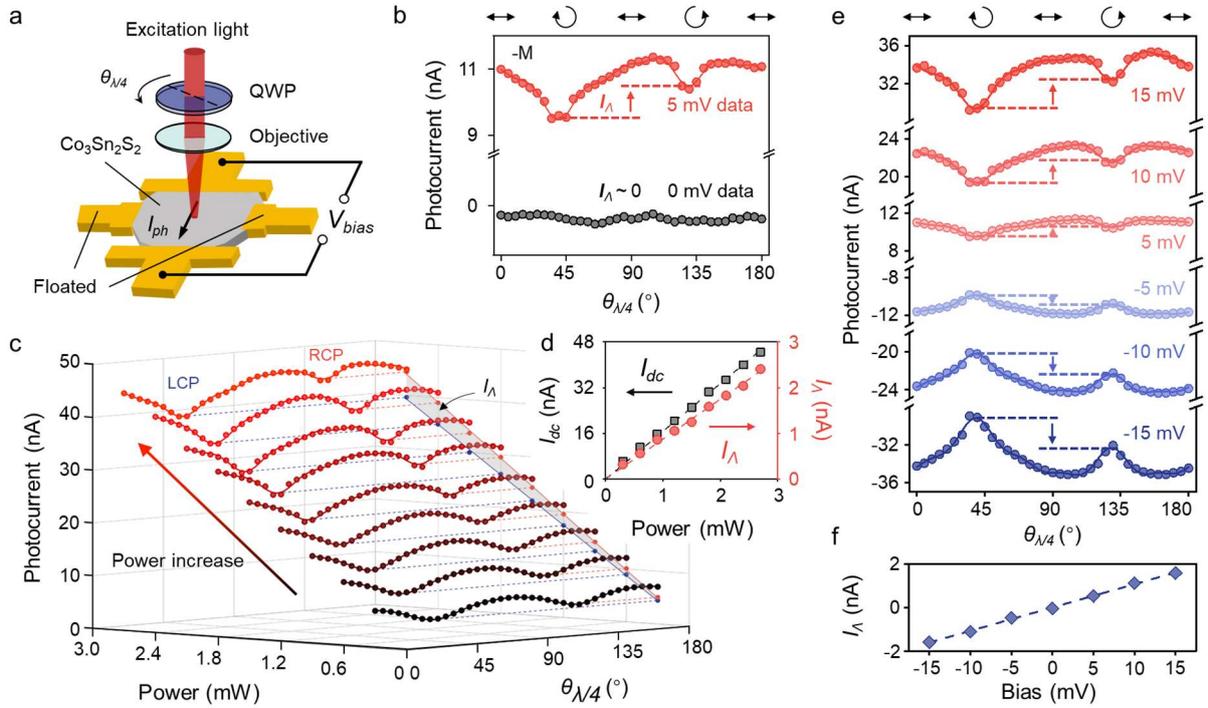

**Figure 2.** Bias- and excitation power-dependent photocurrent measurements of $Co_3Sn_2S_2$. (a) Experimental configuration of the light chirality-dependent photocurrent measurement. (b) Photocurrent measured at different QWP angles with and without a 5-mV external bias applied. (c) Light chirality-dependent photocurrent under different excitation powers with a 10-mV external bias. The blue (red) dashed line indicates the projection of the photocurrent measured under LCP (RCP) excitation. The blue (red) solid line is the linear fit of the photocurrent versus the excitation power measured under LCP (RCP) excitation. (d) Polarization-independent component $I_{dc}$ and light chirality-dependent component $I_A$ as functions of the excitation power. The dashed lines are the linear fits of the data. (e) Photocurrents measured at different QWP angles under different external biases. (f) Dependence of the light chirality-dependent photocurrent on the external bias. The dashed line is the linear fit of the data. All the measurements were taken under -$M$ magnetization. The excitation power was 1.5 mW, except for the power-dependent photocurrent measurements.

Furthermore, we studied the light chirality-dependent photocurrent as a function of the external bias, and the results are shown in Fig. 2e and 2f. The light chirality-dependent photocurrent response increases with increasing applied external bias voltage, as shown in Fig. 2e. To quantify this relationship, we plotted $I_A$ as a function of external bias, as shown in Fig. 2f. Here, $I_A$ is derived from the 180°-periodic component extracted by the Fourier transform of the $\theta_{\lambda/4}$-dependent photocurrent data. As indicated in Fig. 2f, $I_A$ can be tuned by an external bias and clearly linearly depends on the external bias, which is consistent with Equation (1). The same experiment was also performed using a pair of electrodes along the *x*-axis and showed similar behavior (the results are presented in Supplementary Material Section II). This finding indicates that the observed $I_A$ is independent of the orientation between the sample and the electrodes. In addition, since a static electric field can also be built at the interface between the electrodes, a light chirality-dependent photocurrent with similar behavior can also be observed at the interface between the electrodes and $Co_3Sn_2S_2$ (the results are presented in Supplementary Material Section III). To investigate the dependence of the $I_A$ response on the ferromagnetic Weyl phase, we also repeated the experiments at different temperatures. However, no significant $I_A$ response was observed when the temperature was above the $T_C$ (the results are presented in Supplementary Material Section IV). This result suggests that the formation of the ferromagnetic Weyl phase plays a crucial role in the observation of the light chirality-dependent photocurrent.

**Chiral polarized Weyl fermions injection with chiral mid-infrared photon**

The light chirality-dependent photocurrent results presented thus far are all expected according to the symmetry of the materials and are not directly related to the response of the Weyl fermions. To further investigate the response features of the Weyl fermions, light chirality-dependent photocurrent measurements were also carried out at other wavelengths: 10.6 μm, 1550 nm and 800 nm. The measurement results are shown in Fig. 3a (for mid-infrared excitations) and Fig. 3b (for near-infrared excitations). For comparison, the data measured under 4-μm excitation are also shown. The measurements were performed at -*M* magnetization with a 10-mV bias applied. There are two key experimental features: (1) we do not observe a clear $I_A$ under near-infrared excitations (1550-nm and 800-nm excitations). When the bias voltage is increased, we still do not observe a clear $I_A$ under near-infrared excitations (additional data are presented in Supplementary Material Section V). (2) $I_A$ switches signs when the excitation wavelength changes from 4 μm to 10.6 μm. For 10.6-μm excitation, the photocurrent is stronger under LCP excitation than under RCP excitation ($I_A < 0$), whereas for 4-μm excitation, the photocurrent is stronger under RCP excitation than under LCP excitation ($I_A > 0$). The sign switching of $I_A$ under 10.6-μm and 4-μm excitation is further confirmed by the SPCM results and is repeatable on another device and at a lower temperature of 25 K (more data are presented in Supplementary Material Sections VI and VII). In addition, $I_A$ under 10.6-μm excitation is tunable by an external bias voltage, similar to that under 4-μm excitation (Supplementary Material Section VIII).

The above two wavelength-dependent features of $I_A$ cannot be explained by symmetry analysis, as symmetry analysis reveals whether a light chirality-dependent photocurrent is allowed by lattice symmetry; it does not tell either the direction or the amplitude of $I_A$. The fact that $I_A$ is only observed under mid-infrared excitation suggests that the generation of $I_A$ originates from the transitions between the topological bands around the Weyl points under mid-infrared excitation, which cannot be effectively excited by near-infrared excitation. To illustrate this, we plot the band structure along the direction of the line connecting a pair of Weyl points (Fig. 3c) and mark the transitions under excitations of different wavelengths, as illustrated in Fig. 3d. Since the Weyl points are only approximately 60 meV above the Fermi level, the optical transitions of relatively large-energy photons at near-infrared cannot connect the Weyl bands for both the initial and final states, which implies that these optical transitions are dominated by topologically trivial bands. However, for the specific transitions under 10.6-μm and 4-μm excitation, the optical transitions can only occur on different sides of the Weyl cone. For clarity, we also enlarged the band structures near the Weyl point to clearly show the transitions under 10.6-μm and 4-μm excitation, as shown in Fig. 3e. According to the optical selection rule [4], these two transitions require the absorption of photons with opposite chirality: LCP light is preferable for transitions under 10.6-μm excitation, whereas RCP light is preferable for transitions under 4-μm excitation. This optical selection rule is also confirmed by our numerical calculation of the optical dipole matrix element (details are presented in Supplementary Material Section IX). As shown in Fig. 3f, the optical dipole matrix element is larger for RCP excitation than for LCP excitation on the left side of the Weyl point, whereas it is larger for LCP excitation than for RCP excitation on the right side, which is consistent with the optical selection rule shown in Fig. 1a. Here, the diagram illustrated in Fig. 3e shows the transitions within a single Weyl cone. In real materials, Weyl cones with opposite chirality always exist in pairs, according to the no-go theorem [34, 35]. In $Co_3Sn_2S_2$, a pair of Weyl cones with opposite chirality is linked by inversion symmetry. As illustrated in Fig. 3g, in the absence of an external electric field, circularly polarized light will symmetrically excite a pair of Weyl cones with opposite chiralities, but no net photocurrent can be generated since the contributions from the Weyl cones with opposite chiralities cancel each other. Here, we note that the photocurrent discussed in our work refers to the electric current, rather than the chiral current, which can exist even in the absence of an external electric field. However, in the presence of a DC electric field, as in our experiment, the applied electric field can break the inversion symmetry by modifying the band structure [36] and tilting the Fermi surface [37], as illustrated in Fig. 3h and 3i. This leads to asymmetric photoexcitation in a pair of Weyl cones with opposite chirality, and the photocurrent generated by chiral Weyl fermions in the two Weyl cones no longer cancel each other. Consequently, Weyl fermions with specific chiral polarization, which is tunable by the direction of the electric field, can be generated by the excitation of circularly polarized mid-infrared light, and the chiral polarizations are opposite when excited by 10.6-μm and 4-μm excitations with the same circular polarization. The established chiral polarized Weyl fermions are subsequently manifested as the generation of a light chirality-dependent photocurrent. Consequently, this will also lead to a sign switch of $I_A$ at these two different mid-infrared wavelengths, which is consistent with the light chirality-dependent photocurrent experiment results shown in Fig. 3a. Under 1550-nm and 800-nm excitation, since the transition energies are far from the Weyl cones, the chirality-dependent optical transition rules of Weyl bands no longer apply. Therefore, a clear $I_A$ signal is not expected in the photocurrent measurements excited with near-infrared photons, which is consistent with the experimental observations shown in Fig. 3b.

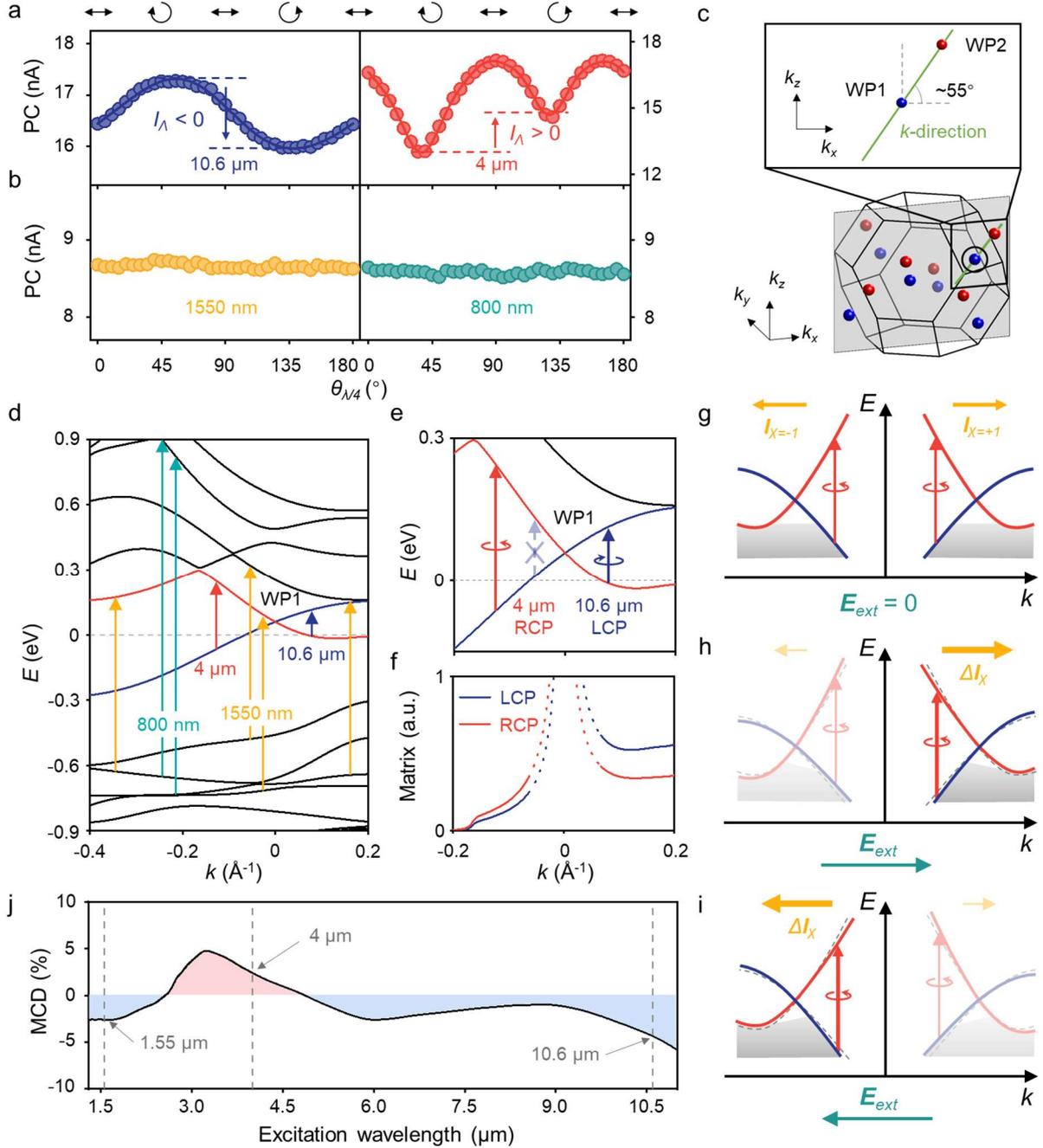

**Figure 3.** Light chirality-dependent photocurrent response at different excitation wavelengths. (a), (b) Light chirality-dependent photocurrent under mid-infrared excitation (a) and near-infrared excitation (b). All the experiments were performed under -$M$ magnetization using the configuration shown in Fig. 2a with a 10-mV bias applied. The excitation powers were 0.9 mW, 0.9 mW, 0.4 mW and 0.6 mW for 10.6-μm, 4-μm, 1550-nm and 800-nm excitations, respectively. (c) Distribution of Weyl points in the Brillouin zone. The green line denotes the momentum cut to draw the band diagram shown in (d). The gray plane represents the $k_x$-$k_z$ plane where the line cut is located. The black circle denotes the Weyl point plotted in (d). Inset: Schematic of the direction of the momentum cut in the $k_x$-$k_z$ plane. The positions ($k_x$, $k_y$, $k_z$) of the pair of Weyl points are at (0.5395, 0, 0.0668) Å$^{-1}$ (WP1) and (0.8110, 0, 0.4578) Å$^{-1}$ (WP2), respectively. The origin of the $k$-direction is set at the position of WP1. (d) Band structure along the direction of the line connecting

a pair of Weyl points marked in (c). The topological bands related to the Weyl point are marked in red and blue. The blue, red, yellow and cyan arrows represent optical transitions induced by 10.6-μm, 4-μm, 1550-nm and 800-nm excitations, respectively. For clarity, we only marked parts of the transitions that are allowed under 1550 nm and 800-nm excitation. (e) Enlarged band structure near the Weyl points. (f) Optical dipole matrix elements for interband transitions between the Weyl bands along the momentum cut in (e). The dashed line indicates the region where optical transitions are forbidden owing to Pauli blocking. (g)-(i) Schematics of optical transitions under the excitation of circularly polarized light in a pair of Weyl cones without an applied electric field (g) and with an applied electric field along the $+k$ direction (h) and $-k$ direction (i), respectively. The dashed lines in (h) and (i) represent the band structure without applying an electric field. The changes in the band structure and tilting of the Fermi surface are exaggerated to clearly illustrate the effect of an electric field that breaks the IS and does not represent the actual changes. (j) Numerical calculations of the MCD spectra under excitation at different wavelengths.

Here, we emphasize that Fig. 3e only presents a qualitative explanation by considering the related optical transition position along the direction of the line connecting a pair of Weyl points. In fact, the transition can also occur at other positions that satisfy the energy conservation and optical selection rules, and the total photocurrent is the sum of contributions from the entire Brillouin zone. Depending on the specific band structure, transitions of 10.6-μm and 4-μm excitations can be simultaneously allowed on the same side of a Weyl cone, which is different from the specific line cut shown in Fig. 3e. In those cases, the transitions of both wavelengths contribute the same sign to the $I_A$ signal, and the total contributions of these parts should be minor according to the experimentally observed sign switch behavior of the $I_A$ signal at 10.6 μm and 4 μm. However, a comprehensive numerical simulation is not possible at the moment, as it requires the calculation of the third-order nonlinear optical conductivity, which is a challenging task because the effects of the DC field on the band structure and the Fermi surface require careful consideration of the scattering process, which is difficult to estimate.

In addition to the process discussed above, the third-order nonlinear photocurrent also contributes to other processes (a detailed discussion is presented in Supplementary Material Section X). In magnetic materials, the major other potential contribution to the $I_A$ arises from magnetic circular dichroism (MCD), which has been studied in various magnetic materials [38−40]. To assess the contribution of the MCD to the observed $I_A$ in $Co_3Sn_2S_2$, we numerically calculated the wavelength dependence of the MCD, which is shown in Fig. 3j. Here, the MCD is defined as $\frac{2(\alpha_{RCP}-\alpha_{LCP})}{\alpha_{RCP}+\alpha_{LCP}}$, where $\alpha_{RCP}$ ($\alpha_{LCP}$) denotes the absorption coefficient for RCP (LCP) light and is calculated using the optical conductivity acquired from experimental results [24] (details are presented in Supplementary Material Section XI). Although the MCD spectra also exhibit opposite signals under 4-μm and 10.6-μm excitations, they show a broadband response with a significant magnitude under 1550-nm excitation, comparable to that under 4-μm and 10.6-μm excitations. However, as shown in Fig. 3b, we do not observe any clear $I_A$ response under 1550-nm excitation, which suggests that the MCD only has a minor contribution to the observed $I_A$ in the mid-infrared region. Therefore, we can conclude that the sign switch of $I_A$ at 4 μm and 10.6 μm is dominated by the coupling between light and Weyl fermions instead of a normal contribution from the MCD effect.

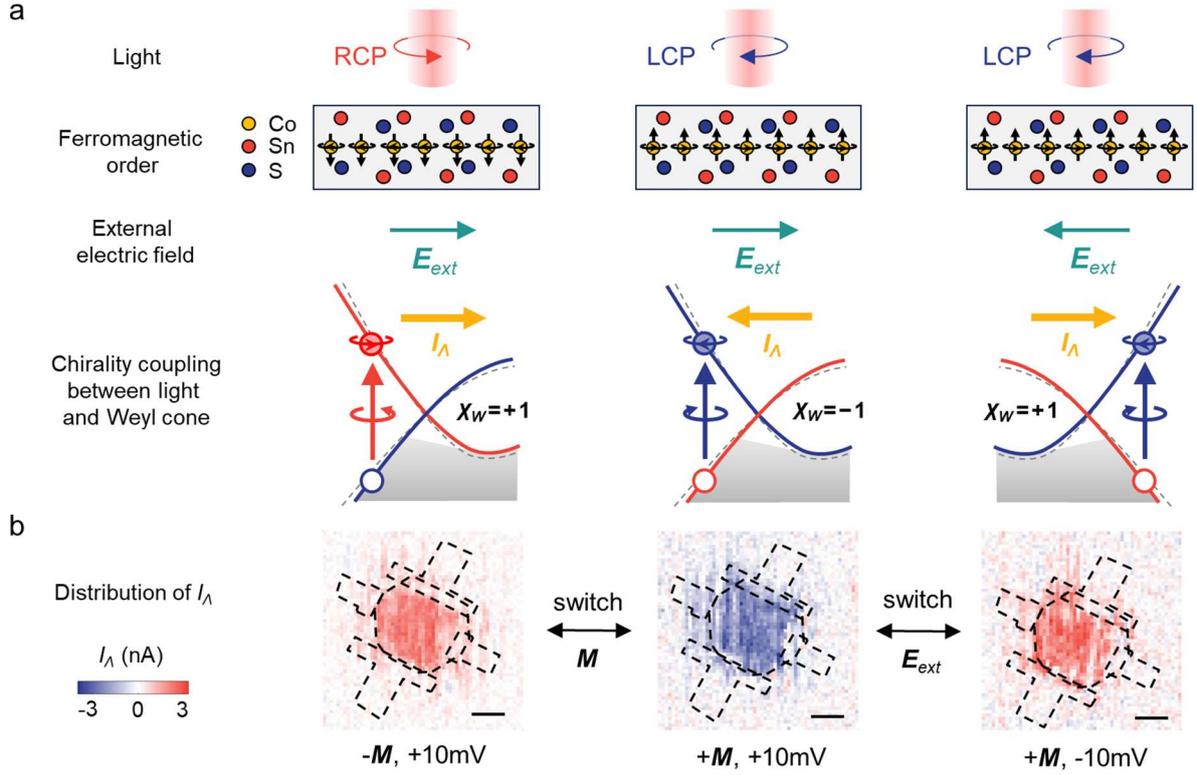

**Figure 4.** Tunable injection of chiral polarized Weyl fermions. (a) Schematic illustration of the tunable injection of chiral polarized Weyl fermions. In the presence of an electric field, the chirality of circularly polarized light can couple with the chirality of magnetic Weyl cones, resulting in the injection of chiral polarized Weyl fermions, which manifests as the generation of $I_A$. The injection of chiral polarized Weyl fermions can be flexibly controlled by the FM order or the external electric field. The dashed lines in the Weyl cone represent the band structure without applying an electric field, similar to those in Fig. 3h and 3i. (b) Distribution of $I_A$ measured under opposite magnetization and electric field directions. All scale bars are 10 μm.

On the basis of the excitation wavelength-dependent features of $I_A$, incorporating the numerical calculation above, the observation of light chirality-dependent photocurrent generation can be attributed to the generation of chiral polarized Weyl fermions. As summarized in Fig. 4a, the chirality coupling of the light with the Weyl cones, in accordance with optical selection rules, can lead to the excitation of Weyl fermions in the Weyl cone. In the presence of an external electric field, which breaks inversion symmetry, the coupling leads to the optical injection of more Weyl Fermions with specific chirality than those with opposite chirality, which helps build an imbalanced population of chiral polarized Weyl fermions, and the consequent effect is detectable through light chirality-dependent photocurrent generation in the photocurrent measurement. Since the light chirality-dependent photocurrent originates from a third-order nonlinear effect, the injected chiral polarization of Weyl fermions can be flexibly tuned by an external electric field. In addition, the magnetism in the TRS-breaking system also provides an additional means to modulate the photoresponse [38, 41, 42] because the chirality of the Weyl cone is directly related to the magnetic order in TRS-breaking WSMs [27]. To demonstrate the tunability of the injected chiral polarization of Weyl fermions, we measured $I_A$ under opposite magnetization and electric field directions, as shown in Fig. 4b. The distribution of $I_A$ is presented to illustrate the global tunable effect across the entire central area of the device, which is measured through the SPCM (details are presented in Supplementary Material Section VI). When either the direction of magnetization or the applied electric field is reversed, the sign of $I_A$ changes accordingly, indicating the controllable optical injection of chiral polarized Weyl fermions. Additional results measured at the central area of the sample are also presented in Supplementary Material Section XII, which demonstrate similar tunability. Practically, the magnetic order can be flexibly controlled by external fields, such as an external magnetic field or an electric field, through spin-transfer torque or spin-orbit torque to switch the magnetic order [43–46] or changing the temperature to induce magnetic phase transitions. Besides, from the perspective of symmetry, the emergence of the $I_A$ response only relies on the breaking of IS, which prevents the cancellation of contributions between a pair of Weyl cones with opposite chirality. Therefore, the application of an electric field to break IS, as demonstrated in our work, represents only one possible approach. This effect could also be observed by employing other means to break IS, such as applying inhomogeneous strain and tailoring appropriate Van der Waals heterointerfaces. Consequently, versatile control through multiple external fields makes TRS-breaking WSMs potentially better material platforms for quantum control of the couplings between different chiral degrees of freedom than IS-breaking WSMs.

## SUMMARY

In summary, the coupling between the chirality of light and magnetic Weyl cones in magnetic WSM $Co_3Sn_2S_2$ provides a convenient route to inject chiral polarized Weyl fermions via chiral polarized mid-infrared photons. As demonstrated experimentally, such injection of chiral polarized Weyl fermions can be flexibly tuned by changing the external electric field or the ferromagnetic order, revealing more versatile tunable chirality couplings between light and magnetic Weyl cones than IS-breaking Weyl semimetals. The chirality of Weyl fermions, which serve as a new degree of freedom, has been proposed to encode and process information, given its ability to couple with the external control field [5]. On the device physics side, the optical injection of chiral polarized Weyl fermions also allows researchers to study the behavior of magnetic Weyl fermions in the excited state through transient optical spectroscopy, including their interaction with various external fields, such as magnetic and electric fields, as well as their ultrafast dynamics. These excited-state studies complement the transport measurements that typically focus on behaviors under equilibrium conditions and are indispensable for understanding the nontrivial topological properties of magnetic Weyl fermions. The highly tunable chirality couplings between light cones and Weyl cones revealed in the present work, together with their exotic transport properties and versatile magnetic control [22, 23, 27, 47], magnetic WSMs not only provide an ideal platform for exploring and manipulating the couplings of chiral degrees of freedom but also promise unprecedented applications in novel optoelectronic, nanophotonic and spintronic devices.

# METHODS

## Sample growth and device fabrication

The chemical vapor transport (CVT) method was used to grow $Co_3Sn_2S_2$ nanoflakes with different thicknesses from 100 to 10 nm. The polycrystalline material was sealed in a quartz tube under high vacuum and used as a precursor. A high-quality single-crystal sample with a good hexagonal morphology was selected to fabricate the device for measurement. Standard electron beam lithography was employed to fabricate the electrode pattern. Ti/Au was deposited via electron beam deposition for electrical contact.

## Magnetization process

The magnetization of the sample in the experiment was achieved through different cooling cycles in the presence of a magnetic field. The magnetic field is provided by a NdFeB magnet, which is placed on the optical window directly above the device. The magnetic field strength is approximately 0.25 T where the device is installed. The magnetization direction is determined by the orientation of the NdFeB magnet.

## Photocurrent measurement

In the photocurrent measurements, continuous-wave light from 10.6 μm, 4 μm, 1550 nm and 800 nm lasers is focused to spot sizes of ∼20, 10, 3 and 2 μm, respectively. The laser beam is modulated via a mechanical chopper (331 Hz), and the photocurrent signal is detected via a current preamplifier (DL Instruments 1211) and a lock-in amplifier (Stanford Research systems SR830). For the polarization-dependent photocurrent measurements, a linear polarizer was used to ensure the linear polarization of the incident light. Quarter wave plates of different wavelengths are rotated electrically to obtain circularly polarized light, and the photocurrent is recorded as a function of the angle of the QWPL ($\theta_{\lambda/4}$). For bias-dependent measurements, an external source-drain bias voltage is applied through an external voltage source.

## Density functional theory (DFT) calculations

The electronic band structures were calculated via the Vienna ab initio simulation package (VASP) with the projector-augmented-wave (PAW) approach [48]. The exchange and correlation energies were considered in the generalized gradient approximation following the Perdew–Burke–Ernzerhof parametrization scheme [49]. We projected the Bloch wavefunctions into maximally localized Wannier functions (MLWFs) and constructed a tight-binding model Hamiltonian based on the overlap of MLWFs [50].

# SUPPLEMENTARY DATA

Supplementary data are available at *NSR* online

**FUNDING**

This project was supported by the National Key Research and Development Program of China (Grant Nos. 2020YFA0308800, 2021YFA1400100, and 2019YFA0704900), the National Natural Science Foundation of China (Grant Nos. 12034001, 62250065, 62325401, 12034003, 52088101, 62227822, and 12274003), the Open Fund of State Key Laboratory of Infrared Physics (Grant No. SITP-NLIST-ZD-2023-02) and the Innovation Program for Quantum Science and Technology (Grant No. 2021ZD0302600).

**AUTHOR CONTRIBUTIONS**

D.S. conceived the idea and designed the research program. J.Y. grew the $Co_3Sn_2S_2$ nanoflakes, fabricated the devices, and performed the transport measurements under the supervision of E.L.; Z.F., J.M., S.C., D.L., D.Y., and C.X. performed the optical measurements under the supervision of A.P. and D.S.; Y.S., Z.L., performed the theoretical calculations under the supervision of J.F. and J.C.; and Z.F., J.C., E.L. and D.S. analyzed the results. All the authors discussed the results. Z.F. and D.S. wrote the paper with input from all the authors. D.S. supervised this project.

*Conflict of interest statement:* None declared